\documentclass[twocolumn,showpacs,a4paper,superscriptaddress,nofootinbib,tightenlines,floats]{revtex4}
\usepackage{bm}
\usepackage{latexsym}
\usepackage{dcolumn}
\usepackage{amsfonts,amssymb}
\usepackage{graphicx,epsfig}
\usepackage{psfrag}
\usepackage{amsmath,amssymb}

\begin{document}

\title{\textbf{Zero-spin-photon hypothesis: `Zero-spin-photon
generation in pair-production and its subsequent decay into neutrino
and antineutrino' - solves many-riddles of physics and universe}}

\author{R. C. Gupta}
\email{rcg_iet@hotmail.com, rcgupta@glaitm.org}
\affiliation{Institute of Technology (GLAITM), Mathura-281 406,
India}

\author{Anirudh Pradhan}
\email{acpradhan@yahoo.com} \affiliation{Department of Mathematics,
Hindu Post-graduate College, Zamania-232 331, Ghazipur, India}

\author{Ruchi Gupta}
\email{ruchig@standfordalumni.org} \affiliation{Cisco Systems, San
Jose, California, USA}

\author{Sanjay Gupta }
\email{sanjaygu@gmail.com} \affiliation{True Demand, Los Gatos,
California, USA}

\author{V. P Gautam}
\email{vpgautam@yahoo.com} \affiliation{Theoretical Physics, IACS,
Calcutta, India}

\author{B. Das}
\email{bdas4569@yahoo.com} \affiliation{Department of Physics,
Lucknow University, Lucknow-226 007, India}

\author{Sushant Gupta}
\email{sushant1586@gmail.com} \affiliation{Department of Physics,
Lucknow University, Lucknow-226 007, India}


\date{\today}

\begin{abstract}
`What is work and what is heat' is re-investigated from the
perspective of second law of thermodynamics. It is shown that the
inevitable consequence of second law of thermodynamics and spin
conservation necessitates the possible generation of zero spin
photon in pair production process, and its subsequent decay explains
the birth of neutrino and antineutrino. The proposed
neutrino-genesis, solves many riddles of physics and universe. The
riddles considered and explained are about: (i) mysterious neutrino
(and antineutrino) and its bizarre properties such as handed-ness
and parity-violation, (ii) questionable asymmetry/ excess of matter
over antimatter, (iii) possibility of existence of antimatter world
and (iv) parity (P) violation and aspects of CP and CPT violation or
restoration in the universe.
\end{abstract}

\maketitle

\section{INTRODUCTION}
Science has progressed a lot, but leaves several unsolved riddles
here and there. Any convincing `Proposal' even at the level of
`hypothesis', which solves some of the riddles, should be welcomed.
The authors propose such a hypothesis: `Zero-spin-photon generation
in pair-production and its subsequent decay into neutrinos and
antineutrinos'.\\\\
The riddles considered here are about:\\
(1)Mysterious neutrinos (and antineutrinos) and its bizarre
properties? \\
(2)Is there a need of slight asymmetry of matter over antimatter (to
explain extremely low value of nucleon to photon ratio) in
early universe?\\
(3)Possibility of existence of anti-matter world?\\
(4)Parity-violation (and also about CP and CPT)?\\\\
These riddles could easily be explained with the `zero- spin-photon
hypothesis' considerations discussed in this paper. The proposed
`zero-spin-photon' is in fact attributed as a necessary out-come in
pair-production process as-demanded-by the
second-law-of-thermodynamics. Moreover, since the zero-spin-photon
(boson with velocity c) does not exist in nature, it must be
unstable to decay into two smaller half-spin particles such as
neutrino and antineutrino (fermions with velocity c) conserving the
spin. The hypothesis about `zero-spin-photon and its subsequent
decay into half-spin neutrino and antineutrino', clearly explains
the genesis and bizarreness of neutrinos and solves several riddles.

\section{ZERO-SPIN-PHOTON GENERATION IN PAIR-PRODUCTION AND ITS
SUBSEQUENT DECAY INTO NEUTRINOS AND ANTINEUTRINOS}

The authors have indicated in an earlier paper \cite{ref1} the
possibility of low-energy zero-spin-photon ($\gamma_{0}$) generation
from high-energy $\gamma$-ray-photon in pair production as $\gamma =
e^{-} + e^{+} + \gamma_{0}$. To explain the essence of the paper
\cite{ref1} , the re-understanding of `heat' and `work' in view of
second-law-of-thermodynamics is necessary, which is explained as
follows in section $II(A)$.

\subsection{Heat and Work}
Second law of thermodynamics implies that `although work can be
fully converted to heat,but heat can-not be fully converted to
work'. For conversion of heat (radiation) to work (particle) some
heat must go as waste. Efficiency of `work to heat conversion' could
be = $100 \%$, but efficiency of `heat to work conversion' must be $
< 100\%$. It may also be noted/remembered that though heat is
considered as a statistical (bulk) aspect, but thermodynamics is
equally applicable even for single-photon interactions, as shown
and discussed in the earlier paper \cite{ref1}.\\\\
In thermodynamic-process `heat' and `work' are generally obvious,
but there are some misconceptions too. The so-called `heat of a hot-
body', as per second law of thermodynamics, is in fact not `heat'
but `work' as it is due to vibration/motion of atoms/molecules. In
electronic process where usually
`energy'-transfer/transition/conversion take place, recognition of
heat and work is even more difficult. What is energy? Is Energy
`heat' or `work'? A little thermodynamics considerations
\cite{ref1}, however, will reveal that all energies such as
potential energy, kinetic energy, electrostatic energy, chemical
energy, nuclear energy, mass energy $m c^{2}$ etc. are in a way
`work', except the radiation energy $h \nu$ which is the real
`heat'. In fact `heat' is the energy carried by the mass-less
particle such as photon radiation waves, whereas energy carried by
massive particle is `work'. In other words, boson (Photon) carries
the `heat' as radiation waves, whereas fermion (electron or
fermion-groups as atoms/molecules) carries `work' as kinetic,
potential energy and other energy of particle(s). With this
understanding that the photon (radiation $h \nu$) is `heat' and that
the particle mass (energy $ mc^{2}$) is work; the two `mass and
energy conversion' process namely - `annihilation' and
`pair-production' [2-4] of electron and positron are reexamined from
thermodynamics point of view, as follows.

\subsection{Annihilation and Pair-Production}
\subsubsection{Annihilation}
A particle(say, electron $e^{-}$) and an antiparticle (say, positron
$e^{+}$) can annihilate each other giving two $\gamma$-ray photons
i.e.,
\begin{equation}
\label{eq1}
 e^{-} + e^{+} = 2\gamma.
\end{equation}
The total mass energy $1.02$ MeV plus kinetic energy of the
particles fully converts $100\%$ to the energy of the photons. The
full conversion of work (mass energy) to heat (radiation) is well
permissible under second law of thermodynamics; the exact-reverse,
however, is not permissible as discussed in next Section $IIB(2)$
and this is the key point of this paper.
\subsubsection{Pair-production and pairs-production}

If a $\gamma$-ray of radiation energy (heat) of $ h \nu = 1.02 $ Mev
(exactly) could have to produce the electron and positron pair
$2mc^{2}$ (work), then it would be $100\%$ conversion of heat to
work against the second law of thermodynamics, thus impossible.  To
save (satisfy) the validity of the second law of thermodynamics, a
higher energy ($ h \nu > 1.02$ MeV) photon is required and that some
other object such as nucleus is involved in the pair-production
process to carry-away part of initial photon-energy (and momentum),
thus only less than $ 100 \%$ photon's energy (heat) is utilized in
producing the `electron
positron' pair $2mc^{2}$(work).\\\\
{\bf{(a) Zero-spin-photon hypothesis}}\\\\
The authors propose the hypothesis that: `if pair-production is ever
to happen in empty space (such as in early universe), then it would
be like $\gamma = e^{-} + e^{+} + \gamma_{0}$ giving a
zero-spin-photon($\gamma_{0}$) of low energy out of the mother
$\gamma$-ray photon of high energy of more than $1.02$ MeV,
in-accordance with second law of thermodynamics and spin
consideration'. The hypothesis of zero-spin-photon as given in Eq.
(\ref{eq2}) is a necessity in pair-production in empty-space to save
(satisfy) the second law of thermodynamics and to conserve spin.
\begin{equation}
\label{eq2} \gamma= e^{-} + e^{+} + \gamma_{0}.
\end{equation}
Note that pair-production simply as $\gamma = e^{-} + e^{+}$
(specially in empty space) is thermodynamically wrong, thus not
possible. There are, however, four or five possibilities of
pair-production to save(satisfy) second-law of thermodynamics, as
follows:
\begin{equation}
\label{eq3} \gamma_{+1} = e^{-}_{+\frac{1}{2}} +
e^{+}_{+\frac{1}{2}} + \gamma_{0}.
\end{equation}
\begin{equation}
\label{eq4} \gamma_{-1} = e^{-}_{-\frac{1}{2}} +
e^{+}_{-\frac{1}{2}} + \gamma_{0}.
\end{equation}
or
\begin{equation}
\label{eq5} \gamma_{+1} = e^{-}_{+\frac{1}{2}} +
e^{+}_{-\frac{1}{2}} + \gamma^{'}_{+1}.
\end{equation}
\begin{equation}
\label{eq6} \gamma_{+1} = e^{-}_{-\frac{1}{2}} +
e^{+}_{+\frac{1}{2}} + \gamma^{'}_{+1}.
\end{equation}
\begin{equation}
\label{eq7} \gamma_{-1} = e^{-}_{+\frac{1}{2}} +
e^{+}_{-\frac{1}{2}} + \gamma^{'}_{-1}.
\end{equation}
\begin{equation}
\label{eq8} \gamma_{-1} = e^{-}_{-\frac{1}{2}} +
e^{+}_{+\frac{1}{2}} + \gamma^{'}_{-1}.
\end{equation}
or
\begin{equation}
\label{eq9} \gamma_{+1} = e^{-}_{-\frac{1}{2}} +
e^{+}_{-\frac{1}{2}} + \gamma^{''}_{+2}.
\end{equation}
\begin{equation}
\label{eq10} \gamma_{-1} = e^{-}_{+\frac{1}{2}} +
e^{+}_{+\frac{1}{2}} + \gamma^{''}_{-2}.
\end{equation}
Depending on the spin considerations; possibilities given in
equations (\ref{eq3}) and (\ref{eq4}) show emission of low energy
zero-spin-photon ($\gamma_{0}$), whereas possibilities given in
equations (\ref{eq5}) to (\ref{eq8}) show emission of usual /normal
(but weaker) spin one photon ($\gamma^{'}$) and that possibilities
in equations (\ref{eq9}) and (\ref{eq10}) indicate the possible
graviton ($\gamma^{''}$) emission. Note that though process are
given in equation (\ref{eq5}) to (\ref{eq10}) are possible (single)
{\bf pair}-production, but not considered any further; whereas the
interesting process given in equation (\ref{eq3}) and (\ref{eq4})
can lead to (double) {\bf pairs}-production after subsequent decay
of the zero-spin-photon $\gamma_{0}$ to neutrino and antineutrino
(Eq. (\ref{eq11})) as shown in Figure $1$ and in Eq. (\ref{eq12}) or
(\ref{eq13}). Combination of the above equations, such as $\gamma =
e^{-} + e^{+} + \gamma_{0} + \gamma^{'}$, are also possible but not
considered further because our concentration is only on Eq.
(\ref{eq2}) or on (\ref{eq3}) and (\ref{eq4}) to emphasize the
importance of $\gamma_{0}$ and its subsequent decay via Eq.
(\ref{eq11}).\\

\begin{figure}[htbp]
\centering
\includegraphics[width=8cm,height=8cm,angle=0]{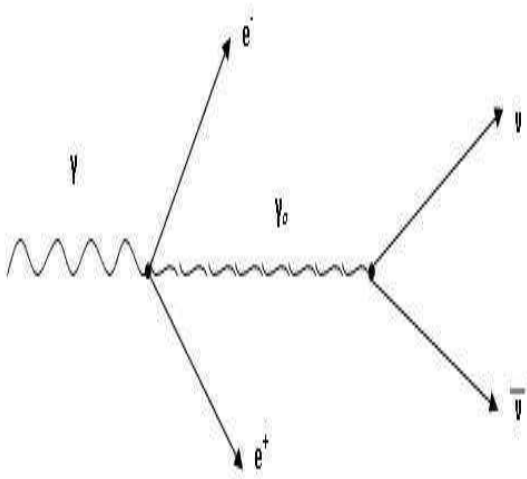}
\caption{Pair-Production and Pairs-Productions}
\end{figure}
{\bf{(b) Subsequent decay of zero-spin-photon to half-spin neutrinos
and
antineutrinos, making the pair-production as pairs-production}}\\\\
Since the bizarre-looking zero-spin-photons (with velocity c) are
not seen to exist in nature, it must have been unstable to decay
into two smaller half-spin particles such as the known neutrino
($\nu$) and antineutrino ($\bar{\nu}$) (with velocity c) as follows.
\begin{equation}
\label{eq11} \gamma_{0} = \nu + \bar{\nu}.
\end{equation}
Combining Eqs. (\ref{eq2}) and (\ref{eq11}), the equation of
pair-production can be written as
\begin{equation}
\label{eq12} \gamma = e^{-} + e^{+} + \nu + \bar{\nu}.
\end{equation}
which shows that not only one-pair ($ e^{-}$ and $ e^{+}$) but
two-pairs ($ e^{-}$ and $ e^{+}$, and $ \nu $ and $\bar{\nu}$) are
produced. The thermodynamically-valid and spin-conserved
pair-production given in (\ref{eq2}) finally becomes
pairs-production (\ref{eq12}), as shown in Figure $1$. The Eq.
(\ref{eq12}), if spin-included, can be re-written as.
\begin{equation}
\label{eq13} \gamma_{\pm1} = (e^{-}_{\pm\frac{1}{2}} + e^{+}_{\pm
\frac{1}{2}}) + (\nu_{-\frac{1}{2}} + \bar{\nu}_{+\frac{1}{2}}).
\end{equation}

\section{THE RIDDLES RESOLVED}

\subsection{Neutrino $ \nu$ (antineutrino $ \bar{\nu}$ ) and its bizarre
 properties such as handedness (helicity) etc.}

\subsubsection{Birth of $\nu$ and $\bar{\nu}$ and its handedness (helicity) }

The most mysterious and elusive of all particles are neutrinos and
antineutrinos \cite{ref5}. However, only the basic and fundamental
aspects of it are considered here. It is not very much clear - why
and how these are produced specially with such left or right
handedness
and with bizarre characteristics.\\\\
It is remarkable to mention here that in Eqs. (\ref{eq3}) and
(\ref{eq4}), the zero-spin-photon $\gamma_{0}$ is one of the extra
outcome of pair-production, necessary to save the second law of
thermodynamics. It is also observed from Eq. (\ref{eq11}) that this
zero-spin-photon (moving with velocity c) subsequently decays into
neutrino and antineutrino (also moving with velocities c in same
direction), the birth of neutrino $\nu$ (and antineutrino
$\bar{\nu}$), is re-written as follows indicating the spins.
\begin{equation}
\label{eq14} \gamma_{0} = \nu_{-\frac{1}{2}} +
\bar{\nu}_{+\frac{1}{2}}.
\end{equation}
The reason why one of the $\nu$ is left-handed (spin
${-\frac{1}{2}}$) and the other ($\bar{\nu}$) is right-handed (spin
${+\frac{1}{2}}$) is explained as follows. As shown in Figure $2$,
$\gamma_{0}$ moves in a direction with velocity c, and when it
decays into neutrinos and antineutrinos both must move in same
direction to conserve momentum and both must have spin-rotation in
opposite direction to conserve spin. Thus to conserve momentum and
spin, one (neutrino) is left-handed (${-\frac{1}{2}}$) and other
(antineutrino) is right-handed (${+\frac{1}{2}}$). This is so- is
known to be true experimentally and is famous as parity-violation
\cite{ref2,ref5,ref6}.

\begin{figure}[htbp]
\centering
\includegraphics[width=8cm,height=8cm,angle=0]{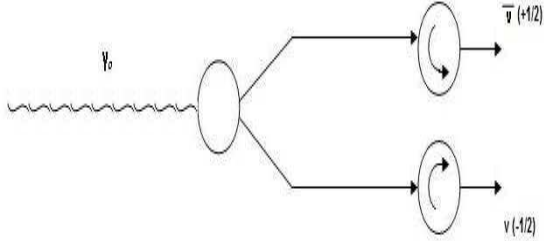}
\caption{Handedness (Helicity) of Neutrinos and Antineutrinos}
\end{figure}

\subsubsection{Nomenclature and characteristics}

Neutrinos (and antineutrinos) are born out of decay of the latent
zero-spin-photon $\gamma_{0}$ as per Eq. (\ref{eq11}) or
(\ref{eq14}). This zero-spin-photon could also be named as
mesonic-photon. Is zero-spin-photon not like `Goldstone-boson'
\cite{ref2} ? It seems, it is so; and if it is so, it should have
linkage with
Higg's mechanism \cite{ref2}. \\\\
The appropriate names of neutrinos and antineutrinos could be
`half-spin-photon' where half-spin implies ${-\frac{1}{2}}$ for
neutrinos and ${+\frac{1}{2}}$ for antineutrinos. The neutrino (and
antineutrino) alias half-spin-photon seems to have mixed properties
/characteristics of photon and fermion. From spin consideration it
is fermion (lepton), as commonly believed. But since it moves with
velocity of light c, it has characteristic of photon. This mixed
dilemma is in-fact not a dilemma but is solution for many bizarre
properties such as `handedness' (helicity) as just explained ( in
Section $IIIA(1)$, Figure $2$). Hence, the neutrino (and
antineutrino) could be named as `photon-with-handedness', or simply
as `fermionic-photon' which indicates that it has both the aspects
($ v = c$ and spin half) of photon and of fermion. It is more of a
photon that fermion, thus appropriate to be named as
`fermionic-photon' with an optional prefix of left or right to show
the handedness as listed in Table-$1$ (along with some other
familiar particles/antiparticles and its characteristics such as
charge, z-component of spin, handedness, possible speed,
stability/decay
etc.).\\\\
Efficiency of `heat to work conversion' in pair-production,
according to second law of thermodynamics, would be less than
$100\%$ but it could be as high as, say $99.9999\%$; thus a
$\gamma$-ray photon (with energy $h \nu > 1.02$  MeV) could produce
a Zero-spin-photon $\gamma_{0}$ (of energy $ \sim 1 $ eV), which
subsequently splits into tiny (of mass $ \sim 0.5   eV/c^{2}$)
particles:
neutrino and antineutrino (Figures $1$ and $2$).\\\\
High penetrating power of the sharp-tiny (but with small mass)
neutrinos and antineutrinos (moving with velocity of light) could be
due to its unidirectional spin, resulting in its deep drilling
action.\\\\

\begin{table*}
{\bf {Table-1 Some particles and its characteristics}} \\
\mbox{ } \\
\begin{ruledtabular}
\begin{tabular}{lccccc}
{\bf Particle(symbol)}&{\bf spin} & {\bf Handedness}& {\bf Speed} & {\bf Category} &
{\bf Decay} \\ \hline \\
Electron  $(e^{-} )$ &  $(\pm\frac{1}{2})$ & {\rm No} &   $0\leq v<c $ &
fe(lep) &     stable \\

Positron ($ e^{+} $) & $(\pm\frac{1}{2})$ &No &   $0\leq v<c $&
fe(lep) &     stable \\

Proton    ( $ p^{+} $) & $(\pm\frac{1}{2})$ &No &   $0\leq v<c $ &
fe(bar)&     stable  \\

Antiproton($ p^{-} $) & $(\pm\frac{1}{2})$ & No &   $0\leq v<c $&
fe(bar) &     stable \\

Neutron( $ n $) & $(\pm\frac{1}{2})$ & No &   $0\leq v<c $&
fe(bar)&  $n=e^{-}+p^{+}+\bar{\nu}$ \\

Antineutron( $ n'$) & $(\pm\frac{1}{2})$ & No &   $0\leq v<c $&
fe(bar)&     $n=e^{+}+p^{-}+\nu $ \\

Photon     ($ \gamma $) & $(\pm 1) $ & No &   $ v = c $  &
bos(ph)&     stable \\

Zero-spin-ph( $ \gamma_{0}$) & $(0)$ & No &   $ v = c $  &
mes-ph &     $\gamma_{0}=\nu+\bar{\nu}$ \\

Neutrino    ( $ \nu$) &  $(-\frac{1}{2})$ & L-handed &   $ v = c $  &
L-fe-ph&     stable\\

Antineutrino($ \bar{\nu}$) & $(+\frac{1}{2})$ &R-handed &   $ v = c $  &
R-fe-ph&     stable \\
\end{tabular}
\end{ruledtabular} \\ \mbox{ } \\
\footnotetext{} {Here, L, R, ph, fe, bos, lep, mes and bar stand for
Left, Right, photon, fermion, boson, lepton, meson and baryon
whereas mes-ph and fe-ph are understood to be read as mesonic-photon
and fermionic-photon respectively.}
\end{table*}

\subsection{Is there a need of slight asymmetry (excess) of matter
over antimatter (to explain extremely low value of nucleon to photon
ratio $ 10^{-9}$ ) in the early universe?}
{\bf { Present approach:}}

It is well known that our universe contains much more number of
photons ($ \sim 10^{9} $) for each nucleon. This is explained in the
past and contemporary physics \cite{ref3,ref4,ref7,ref8,ref9} as
follows. It is assumed that during particles-production in the early
universe there was slight asymmetry (excess) of matter over
antimatter i.e., more matter particles (e.g., one billion plus one)
were produced as compared to antimatter particles (one billion). The
one billion of each matter and antimatter particles annihilated to
produce two billion photons leaving behind one matter particle, in
each turn. Thus the slight excess of matter particles survived
ultimately giving rise to our matter universe in which nucleon to
photon ratio is $10^{-9}$.\\
{\bf {Alternative approach:}}

Another explanation, however, is possible without the need of any
asymmetry of matter over antimatter, is given as follows. For it, it
is considered that due to symmetry, in early universe exactly same
amount of matter and antimatter were produced (e.g.,one billion plus
one, for each). The one billion each of matter and anti-matter
annihilated to produce two billion photons. Out of these, one
billion photons with one matter particle every time went to create
matter world; whereas one billion photon with one antimatter
particle every time went to produce antimatter world. Thus there
seems a possibility of creation of antimatter world.

\subsection{Possibility of creation, separation and existence of
antimatter world}

{\bf {Interpretation of Fitch-Cronin experiment of Kayon-decay in
both ways: as evidence against and in-favour-of existence of
antimatter-world}} \\\\
The present model of Big-Bang assumes that there was a slight excess
of matter (asymmetry-hypothesis) over antimatter in early universe
perhaps $1$ part in $ 10^9 $. The alternative possibility (as
explained in the previous Section $III(B)$) based on
symmetry-hypothesis is that: matter and antimatter were produced in
exact-equal quantity, but later some how antimatter-world separated
from the matter-world. There is no (not enough) evidence either in
favour or against the possibility of symmetry-hypothesis. There is,
however, the key evidence (Fitch-Cronin experiment, as reported in
\cite{ref2,ref10} in favour of asymmetry-hypothesis is decay of
neutral kaon; but that is particularly not a strong evidence and
that the same kaon decay can be reinterpreted otherwise in favour of
symmetry-hypothesis as well, explained as follows.\\\\
{\bf { Fitch-Cronin experiment of kaon decay: totality in
both-worlds rather supports possibility of antimatter-world:}}\\\\
In our matter-world (as in Fitch-Cronin experiment \cite{ref2,ref10}
the long-lived neutral kaon $k_{L}$ can decay into (i) $ \pi^{+} +
e^{-} + \bar{\nu}_{e} $ or (ii) $ \pi^{-} + e^{+} + \nu_{e} $. For
supporting symmetry-hypothesis both (i) and (ii) should decay with
equal probability, but it is found (experiment in our matter-word)
that probability of mode (ii) is very slightly higher, indicating a
possible excess of matter in our matter-world. It can be argued that
if Fitch-Cronin experiment is done in antimatter-world, the neutral
$ k_{L} $ would decay but therein probability in mode (i) would be
very slightly higher, indicating a possible excess of antimatter in
antimatter-world. Considering both the matter-world and
antimatter-world in totality, the whole universe would be perfectly
symmetrical with equal amount of matter and antimatter. Thus it can
be further argued that the decay of kaon, is not against
antimatter-world, but rather is in its favor if both the matter and
antimatter worlds are considered in totality.\\\\
{\bf {Possible creation of both matter and antimatter worlds and its
separation}}\\\\
In this and next paragraphs it is briefly discussed that how matter
and antimatter worlds were formed, and how it separated out. After
the big bang for each $ 10^{9} + 1 $ matter-particles, $ 10^{9} + 1
$ antimatter particles were generated. The $ 10^{9} $ each from
matter and antimatter annihilated to form photons. The remainder
small number of matter and antimatter particles formed the
corresponding two worlds. Besides large number of photons, matter
world mainly consist of proton $ p^{+} $, electron $ e^{-} $, and
neutron ($ n = p^{+} + e^{-} + \bar{\nu}$); whereas antimatter world
would consist of antiproton $ p^{-} $, positron $ e^{+} $ and
antineutron $ n' $ ($ n' = p^{-} + e^{+} + \nu $). There would be
large number of neutrinos and antineutrinos around. But all the free
neutrinos $ \nu $ would go along with matter sensing (from neutron
decay) that as - if neutron contains its counterpart antineutrinos $
\bar{\nu} $. Similarly all the free antineutrinos would go along
with antimatter. That is why our matter-world contains free abundant
neutrinos (all left-handed); antineutrino comes out only in
neutron-decay or in such weak interactions indicating
parity-violation. Similarly, antimatter world would contain free
abundant antineutrinos; neutrino would come out there as in
antineutron-decay indicating
parity-violation there too.\\\\
Many physicists don't deny and some even agree to the possibility of
existence of antimatter world but wonder why and how it got
separated from our world. Why matter-world and antimatter-world
separated out is difficult to explain because the reason is not
known. But, possibly, it can be explained as follows. An alternative
novel gravity-theory recently proposed by Gupta \cite{ref11}
explains: how `gravity is the residual net electrostatic
attraction/repulsion (due to second-order relativistic effect)
between the charged-constituent of otherwise-neutral matter-atoms'.
Interestingly, it is shown in the paper \cite{ref11} that Newton's
gravitational formula is derivable from Coulomb's formula if applied
between the constituents of atoms of two bodies and if special
relativity considerations are taken into account. This gravity
theory \cite{ref11} also suggests (predicts) that though elementary
particles and antiparticles attract each other but atoms (of
matter-world) and anti-atoms (atoms of antimatter-world) would repel
each other. This repulsion between the matter and antimatter could
be the cause of separation of antimatter-world from the matter-world! \\\\
From Einstein's gravity theory (General Relativity)-point of view
also, this repulsion of matter from antimatter seems possible, if
interpreted as follows. Matter creates a concave dimple (valley)
around it in space-time fabric. The test-matter-mass around this
valley tends to fall into the valley, creating an apparent
gravitational attraction. To the test-antimatter-mass, residing on
the other side of the fabric (horizon), the dimple (valley) would
appear from the other side as a mole (hill); hence the
test-antimatter-mass will fall away from the cliff, creating an
apparent repulsion between matter and antimatter.\\\\
\subsection{Parity violation and also about CP and CPT}
In Section $IIIA(1)$ it is explained how neutrinos (and
antineutrinos) are born due to decay of the zero-spin-photon as $
\gamma_{0} = \nu_{-\frac{1}{2}} + \bar{\nu}_{+\frac{1}{2}} $. Also
explained there, is the fact that why neutrinos are left handed ($
{-\frac{1}{2}} $) and antineutrinos are right handed ($
{+\frac{1}{2}} $), both move in same direction with spin in
different directions to conserve spin,$ {+\frac{1}{2} +
-\frac{1}{2}} = 0 $. Thus all neutrinos (free and abundant in our
matter world) are left handed and all antineutrinos (free and
abundant there in antimatter world) are right handed. This is full
(maximal) parity violation (in weak interaction) as observed
\cite{ref2,ref5,ref6} experimentally too.\\\\

Parity (P), and charge-conjugation (C) taken together, CP seems well
restored initially. But from Fitch-Cronin experiment
\cite{ref2,ref10} of kaon decay, a very small CP violation does seem
to occur in matter world thus favoring small asymmetry as mentioned
in Section $IIIC$. But in the same section, it is also argued that
similar violation (in other way) could occur in antimatter world too
again favoring small asymmetry but in opposite way. Thus in totality
(for both worlds considered together) the symmetry seems to be
restored/retained.\\

For time-reversal symmetry (T) there are compelling reasons that
time-reversal can not be a perfect symmetry, especially in view of
thermodynamic-irreversibility (or second law of thermodynamics)
creating thermodynamic arrow of time \cite{ref8}. Thus T must be
violated \cite{ref12} even though it may be very slight, even
un-noticeable. The slight violation of CP in each world and slight
violation of T together; could yield TCP restoration in each world
and in totality too. The `beauty' is that the two slight asymmetries
(CP and T violations) make an exact symmetry (TCP restoration).
Nature prefers symmetry in the whole universe. The suggested way of
TCP restoration is well in accordance with the `beautiful' TCP
theorem \cite{ref13} which states that `combined effect of T, C and
P (in any order) is an exact symmetry of any interaction'. Brief
description for quick look of violation/Restoration of P, CP, T and
CPT are tabulated in Table-$2$ (at appropriate page).\\
\begin{table*}
{\bf {Table-2 Violation/Restoration of P, CP, T and CPT}} \\
\mbox{ } \\
\begin{ruledtabular}
\begin{tabular}{llll}
 \mbox{ } & Matter World &  Antimatter World &  Whole universe \\
\mbox{ }&\mbox{ } & \mbox{ } &   containing both  \\
\mbox{ }&\mbox{ } & \mbox{ } & matter and antimatter worlds \\
\hline \\
(P)&Maximally violated. & Maximally violated. &Parity violation necessary, \\
\mbox{ } & All neutrinos are left, & All neutrinos are right & since the left handed neutrinos and \\
\mbox{ } & free and abundant. & handed, free and abundant. & right handed antineutrinos are born\\
 & & & out of decay of zero spin-photons.\\
& & & \\
(CP)& Nearly restored but    &  Nearly restoration but &  Symmetry of matter and \\
    & very slight violation  &  very slight violation to & antimatter in totality, with \\
    & suggests asymmetry(slight excess) & to suggest asymmetry & slight violation of CP but \\
    & of matter over antimatter. & (slight excess) of matter & differently in both worlds. \\
    &                            & over antimatter.  &  \\
& & & \\
(T)&  Violation(may be &  Violation(may be slight) & Violation(may be slight) \\
   &  slight) due to thermodynamic & due to thermodynamic  & due to thermodynamic\\
   &  arrow of time. & arrow of time.  & arrow of time.\\
& & & \\
(CPT)&             Restored. & Restored. & Restored. \\

\end{tabular}
\end{ruledtabular}
\end{table*}

\section{DISCUSSION}

{\bf{Thermodynamics and Relativity Linked}}\\\\
The generation of zero-spin-photon $ \gamma_{0} $ in pair-production
is the inevitable consequence of second law of thermodynamics and
spin-conservation, as mentioned in the Section $IIB2(a)$. It is also
interesting to note, as mentioned in the other paper \cite{ref1},
that diverse phenomena such as `second law of thermodynamics
(heat-to-work ratio, efficiency $\frac{W}{Q}_{1} = \eta \leq 1 $)'
and `special-relativity (velocity ratio $\frac {v}{c} = \beta < 1
$)' are intimately linked or in other words are two faces of the
same coin. Subsequent decay of zero-spin-photon $\gamma_{0}$ (moving
with c) into neutrino and antineutrino as half-spin-photon (both
moving with c in the same direction) is discussed in the Section
$IIB2(b)$. The half-spin-photons (fermionic-photons) have dual
property of photons ($v = c$) and of fermions (half spin). It seems
that for both: zero-spin-photon and half-spin-photon, the key is
relativity (or thermodynamics) along with spin-conservation.\\\\
{\bf {Super-symmetry: A new possibility of self-super-symmetry}}\\\\
The half-spin-photons alias fermionic-photons are commonly known as
neutrinos and antineutrinos. If the fermionic-photon is thought
(renamed) as fermionic-boson, its dual nature(and name) reminds of
super-symmetry. Super-symmetry \cite{ref9} is a novel concept, which
states that every fermion could have its bosonic partner and vice
versa and is an important ingredient of super-string theory. Is the
neutrino as fermionic-boson is super-symmetric partner of itself,
and the antineutrino too as fermionic-boson is a super-symmetric
partner of itself? It opens-up a new possibility of a particle being
self-super-symmetric.\\\\
{\bf{ $ E = mc^{2} $ re-interpreted}}\\\\
Re-consideration of energy as work or heat as discussed in Section
$II(A)$ indicates that mass energy $ mc^{2} $ is like work and
radiation energy $\gamma$ is heat. The famous mass-energy equation $
E= mc^{2} $ needs to be re-expressed for annihilation and for
pair-production differently (as follows) in the light of the second
law of thermodynamics validity. For annihilation it is as $ mc^{2} =
\gamma $, but for pair-production (Eqs. (\ref{eq3})- (\ref{eq8})) it
should be as $ \gamma = mc^{2} + \gamma^{'} $ or $ \gamma = mc^{2} +
\gamma_{0} $, where $ \gamma $ and $ \gamma^{'} $ are normal
spin-one high and low energy photons and $ \gamma_{0} $ is the low
energy zero-spin-photon.\\\\
{\bf {Possible by-products in pair(s)-production}}\\\\
The input ingredient for pair-production is the high energy $ \gamma
$-ray photon whereas the output products are electron and positron
with some possible by-products such as listed herein: (i) low energy
normal spin-one photon $ \gamma^{'} $ (see Eqs.
(\ref{eq5})-(\ref{eq8})), (ii) low energy novel zero-spin photon $
\gamma_{0} $ (mesonic-photon) as in Eqs. (\ref{eq3}) and
(\ref{eq4}), (iii) the $ \gamma_{0} $ decays/splits (Eqs.
(\ref{eq11}) and (\ref{eq14}) and Figures $1$ and $2$) into
half-spin photons (fermionic-photons) or i.e., into neutrino and
antineutrino ($ \nu $ and $ \bar{\nu}$) and (iv) bi-spin-photon ($
\gamma^{''} $ in Eqs. (\ref{eq9}) and (\ref{eq10}), possibly
as graviton!).\\\\
{\bf {Neutrinos (and antineutrinos): the three known brands and
other
possible brands}}\\\\
Equations(\ref{eq11}) and (\ref{eq14}) and Figures $1$ and $2$ show
how neutrinos and antineutrinos are created. The zero-spin-photon $
\gamma_{0} $ is born-out in pair-production. The $ \gamma_{0} $ is,
as if, also like a pair of Siamese-twins(of $ \nu $ and $ \bar{\nu}
$) joined at hip, which ($ \gamma_{0} $) subsequently decays
(splits) into two separate particles $ \nu $ and $ \bar{\nu}$.
Before the split the rotation (spin) is restricted to zero but after
the split these two rotate (spin) in opposite way (see, Figure $2$)
while moving in same direction with same velocity $c$. The $\nu$ and
$\bar{\nu}$ therein (Section $IIB(2)$) have been considered, mainly
in reference to understand how `electron-type' neutrinos and
antineutrinos are born. The same concept may, similarly, be extended
for generation of `muon-type' and `tau-type' neutrinos and
antineutrinos too, but correspondingly higher energy
level would be required.\\\\
Furthermore, it appears that similarly as per Eq. (\ref{eq12}),
other brands of neutrinos (and antineutrinos) such as, say,
proton-type or meson-type could also be born. But no such brand of
it are known. The reasons for its absence could be that: (i) no one
has cared-for/noticed it, (ii) these probably unstable brands of
neutrinos (and antineutrinos) may have decayed into the three known
stable neutrinos (and antineutrinos), (iii) its corresponding
zero-spin-photon never decayed to produce such brands and remained
quiet, thus these undecayed zero-spin-photons could be still around
us unnoticed but may account for mysterious \cite{ref9,ref14} dark
matter (or part of it).\\\\
{\bf {Heavier particles synthesis: Elementary or Composite?}}\\\\
The present paper, though capable of explaining many riddles, seems
speculative at several places. The other speculative possibility is
that neutrons($ e^{-} + p^{+} + \bar{\nu} = n $) and antineutrons ($
e^{+} + p^{-} + \nu = \bar{\nu} $), are not elementary particles but
composite particles (as indicated in the brackets), and are created
only after the various pair-production processes have created its
constituent-ingredients. Similarly, other particles such as pions ($
\mu^{-} + \bar{\nu}_{\mu} = \pi^{-} $, $ \mu^{+} + \nu_{\mu} =
\pi^{+} $) could possibly also be composite particles. If it is so
that the heavy particles are composite ones, the primordial
nucleo-synthesis have to be re-investigated !\\\\
{\bf {Antimatter World}}\\\\
The possibility of existence of antimatter-world and its separation
from matter-world is discussed in some details (in Sections $III(B)$
and $III(C)$) with consideration of equal amount of matter and
antimatter without any need of asymmetry in it. Possible existence
of the two worlds (mirror universe) may seem speculative, but is
less speculative than that for the much-advocated multi-universe or
parallel-universes and warped passages \cite{ref15}.\\\\

{\bf {P, CP, T and CPT violation/restoration}}\\\\
As mentioned (in section $III(D)$ and in Table-$2$); parity(P) is
necessarily maximally violated in both matter and antimatter worlds.
Parity with charge conjugation (CP) is very slightly violated but
differently in both worlds. Thermodynamically essential slight
violation of time-reversal (T) indicates thermodynamic arrow of
time. In totality, as well as in each world separately, CPT is
always restored (in agreement to CPT theorem \cite{ref13}).\\\\
The only real asymmetry, separately in both worlds and in totality
as well, is thermodynamic irreversibility (violation of
time-reversal) which is the origin of thermodynamic arrow of time
\cite{ref8}. Without it there would be no past, present or future,
hence nothing would exist or can be perceived. Thus, the key to
our-existence (or anthropic principle \cite{ref16} ) is the `second
law of thermodynamics'.

\section{CONCLUSIONS}

Second law of thermodynamics seems to be the key to many secretes of
physics and universe. As shown, the second law of thermodynamics
(with spin conservation)`demands' generation of a low energy
zero-spin-photon $ \gamma_{0} $ in pair-production as $ \gamma =
e^{-} + e^{+} + \gamma_{0} $. This $ \gamma_{0} $, being unstable,
decays/splits into neutrino and antineutrino. This neutrino-genesis
not only solves many mysteries of neutrinos and antineutrinos but
also explains a few riddles of physics and universe. The novel
concept of `generation of zero-spin-photons in pair-production and
its subsequent decay into neutrinos and antineutrinos' could open
new avenues and vistas for search of truth in wide fields ranging
from particle-physics \cite{ref17} to astro-physics and cosmology
\cite{ref18}.

\acknowledgments Two of the authors (A. Pradhan and Sushant Gupta)
thank the Inter-University Centre for Astronomy and Astrophysics,
Pune, India for providing facility where part of this work was
carried out. The author (R. C. Gupta) thanks IET/UPTU, Lucknow and
GLAITM, Mathura for providing facility for the research. Authors
thank Prof. Kanti Jotania for  his help in formatting tables.

\end{document}